# Advances in Machine Learning for Directed Evolution: A Five-Year Retrospective

**Bruce J. Wittmann**
Microsoft, Office of the Chief Scientific Officer
bwittmann@microsoft.com

The last five-plus years have seen many protein engineering disciplines transformed by advances in machine learning (ML), but the same cannot be said for directed evolution. Reflecting on a previously co-authored perspective, I discuss why I believe this to be the case, arguing that a disconnect between the goals of machine-learning-assisted directed evolution (MLDE) researchers—"identify an optimal protein"—and the goals of directed evolution more broadly—"identify a sufficient protein *given time and resource constraints*"—is a principal culprit. As an example, I highlight how nearly all current MLDE methods neglect to account for the cost of DNA synthesis, resulting in strategies that have limited practical applicability regardless of the underlying models' capabilities. I close by discussing recent works that are exceptions to this overarching trend, and emphasize that the last five years of efforts in ML-assisted protein engineering and the prescribed reframe of MLDE objectives need not be mutually exclusive.

## Introduction

Until recently, to all but those actively involved in its research, "AI" was a concept that was largely limited to the realm of science fiction. But now, fueled by high-profile technological advances,[1,2] extraordinary amounts of capital and research investment,[3] and, less optimistically, fears regarding the implications of these developments,[4–7] it has become a household term, permeating family dinnertime conversation and policy discussions alike.

While it may not dominate the news cycle like the latest advances in chatbots or image/video generation, AI-for-biology has experienced a similar renaissance in near lockstep. Pivotal technological advances,[8–10] a swell of AI-for-biology investment,[11] and a growing concern over the potential security implications of our expanding command of biology[12–14] are all signs of a rapidly evolving field. As anyone attempting to keep current on developments in AI-for-biology will know, the pace of research is breakneck, "conventional wisdom" is transient, and distinguishing durable breakthroughs from hype is a persistent challenge.

Looking back, it is astounding to see how much has changed in such a short period of time. Just five years ago, I co-authored a perspective outlining the promise of machine learning (ML)—a term now effectively supplanted by the more general "AI"—for application in directed evolution (DE) and protein engineering.[15] At the time, I was optimistic about the role that discriminative models, semi-supervised learning, and more traditional active learning strategies would soon play, while taking a comparatively cautious stance on the emerging field of generative biology. In the years since that perspective was published, however, efforts to improve discriminative models for machine-learning-assisted directed evolution (MLDE) have largely fallen to the wayside, with no notable change in modeling performance,[16,17] and semi-supervised learning has all but disappeared from the MLDE vernacular, with its utility for protein fitness prediction increasingly drawn into question.[16–26] Meanwhile, generative modeling has become *the* topic, making headlines for its value in producing biological entities ranging from custom antibodies, to de-novo-designed enzymes, all the way to viable genomes.[27–29]

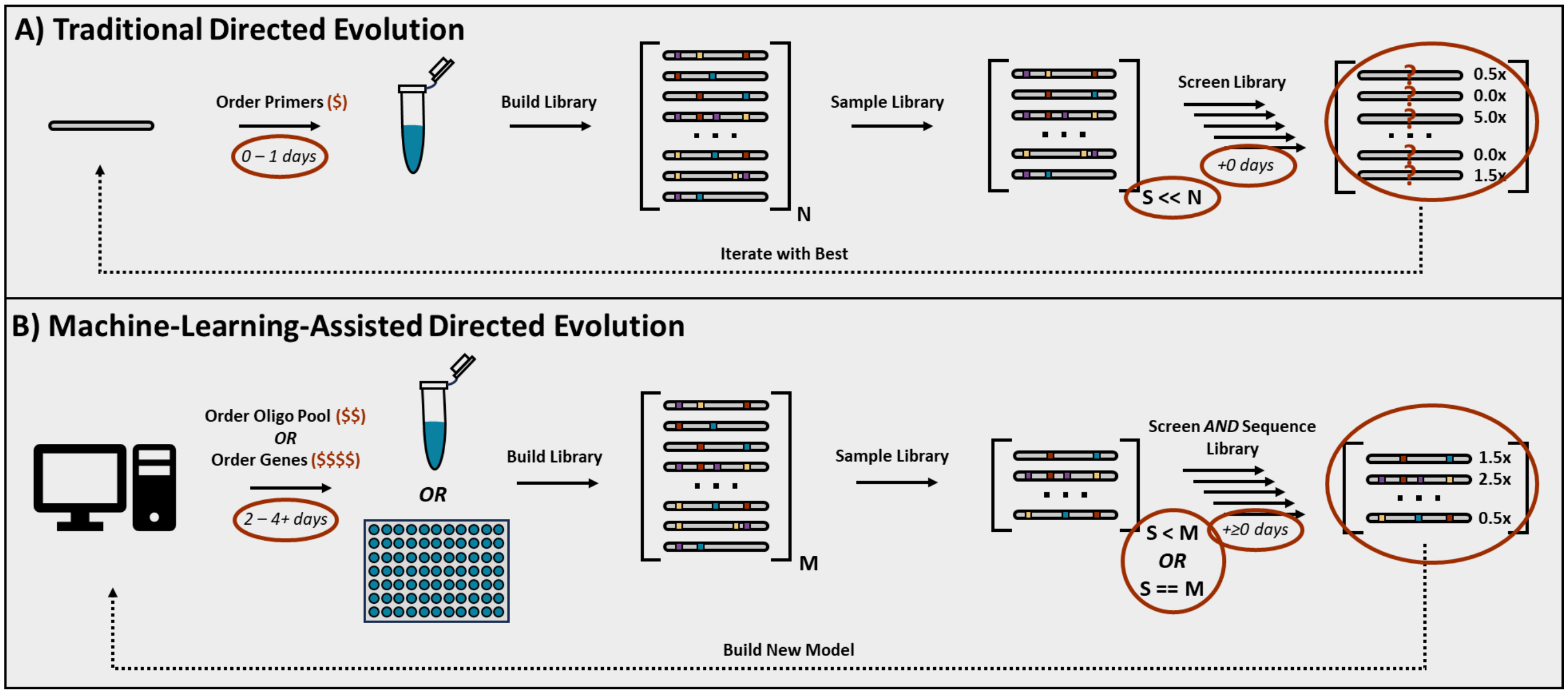

**Figure 1.** A high-level overview of a standard (A) directed evolution (DE) pipeline and (B) machine-learning-assisted directed evolution (MLDE) pipeline, with core differences impacting time and cost requirements of each method circled or bolded in red text. Note specifically that synthesis of oligonucleotides is required by both workflows, but that the types of oligos typically required for MLDE workflows take longer to produce and are far more expensive (based on estimates from Twist Bioscience and IDT as of May 28, 2026). This is a direct consequence of MLDE methods' tendency to prescribe a specific set of variants that should be analyzed per iteration (indicated by the number of variants screened (S) being close to if not equivalent to the library size in MLDE (M) versus being much smaller than the library size in traditional DE (N)). Note also that, in contrast to DE, MLDE methods tend to require that library variants be sequenced. This adds further cost and time to each iteration, though they can be reduced to near-negligible levels by recently developed technologies and workflows.

Perhaps the most striking divergence of reality from prediction, however, is that despite a plethora of paradigm-shifting innovations in AI-for-biology writ large, the same cannot be said for MLDE. While the last five years have seen other disciplines of protein engineering utterly transformed by advances in ML—for example, protein design by models like ProteinMPNN and RFDiffusion,[8,9] or structural biology by models like AlphaFold and RosettaFold[10,30]—the typical DE pipeline of today remains more or less the same.

In what follows, I explain why I believe this to be the case, and what would have to change for the picture to look different five years from now.

## Oft-Ignored Objectives: Cost and Time

Put succinctly, I do not believe that the limited progress in MLDE is the result of failure on behalf of MLDE researchers; rather, I believe it results from a disconnect between many of those researchers' objectives and the fundamental objectives of DE.

DE is an applied technology. Contrary to how it is often depicted, its goal is not to identify the "best" protein, nor even necessarily to identify a locally optimal one; it is to identify a protein that is *good enough given time and resource constraints*. Despite this, evaluations for MLDE methodologies tend to focus primarily on the predictive capabilities of those methods' ML components, entirely ignoring these systems-level constraints.[16,17,31] Were MLDE and traditional DE workflows distinguished only by the former's use of predictive models, such narrowly scoped evaluations may be appropriate. However, as is highlighted in Figure 1, this is not the case: Compared to traditional DE, MLDE (1) relies on more complicated library synthesis strategies and (2) requires a DNA sequencing step, both of which add time and cost to the workflow.

MLDE's reliance on sequencing has long been recognized as a disadvantage, yet continued advances in general sequencing technology, the development of high-throughput assays that directly couple the sequencing and fitness-measurement workflows,[32–34] and the introduction of purpose-built sequencing-for-MLDE strategies have all pushed the added cost of this step (both in time and other resources) toward near-negligible levels.[35,36] The same cannot be said for library synthesis, however.

Whereas traditional approaches can and typically do rely on random mutagenesis strategies for library creation—mostly single- or few-pot reactions requiring only that a few (if any) low-cost oligonucleotide primers be ordered—MLDE-based strategies generally require the synthesis of specific, targeted designs. In the worst case, this means ordering gene-length fragments individually—days to weeks of turnaround time and thousands or more dollars for a moderately sized library. In the best case—barring exceptions for small combinatorial libraries[37]—this means ordering an oligo pool, which might be faster and cheaper than arrayed gene synthesis, but is still considerably more expensive and slower than random-mutagenesis-based strategies.

Figure 2 shows some specific examples that demonstrate this difference in cost. Even assuming the longest (i.e., costliest) standard-length primers and shortest (i.e., cheapest) oligos in an oligo pool, and neglecting the potentially significant added cost imparted by the longer lead times needed for ordering oligo pools and genes, 14 primers can be ordered before any oligo pool is competitively priced (Fig 2A). These 14 primers could be used to build at least 7 error-prone PCR or site-saturation mutagenesis libraries, the latter of which can be used to produce at least 133 variants (far more if combinatorial saturation mutagenesis is employed) and the former of which can be considered (within reason) an effectively boundless source of diversity. Indeed, using error-prone PCR, just 2 primers (flanking the 5' and 3' ends of the region to be engineered) are enough to complete an *entire* engineering campaign—14 is far more than necessary. The comparably priced oligo pool, by contrast, can contain at most 100 120-nucleotide-long variants. For genes, it is worse: only 11 500-nucleotide genes can be ordered for the cost of 14 primers (Fig 2B). For more realistic primer, oligo, and gene lengths, it is worse still: 74 30-nucleotide primers can be ordered before a 100-member oligo pool of 300-nucleotide oligos is competitively priced (Fig 2C); only 9 1500-nucleotide genes could be ordered for similar cost (Fig 2D).

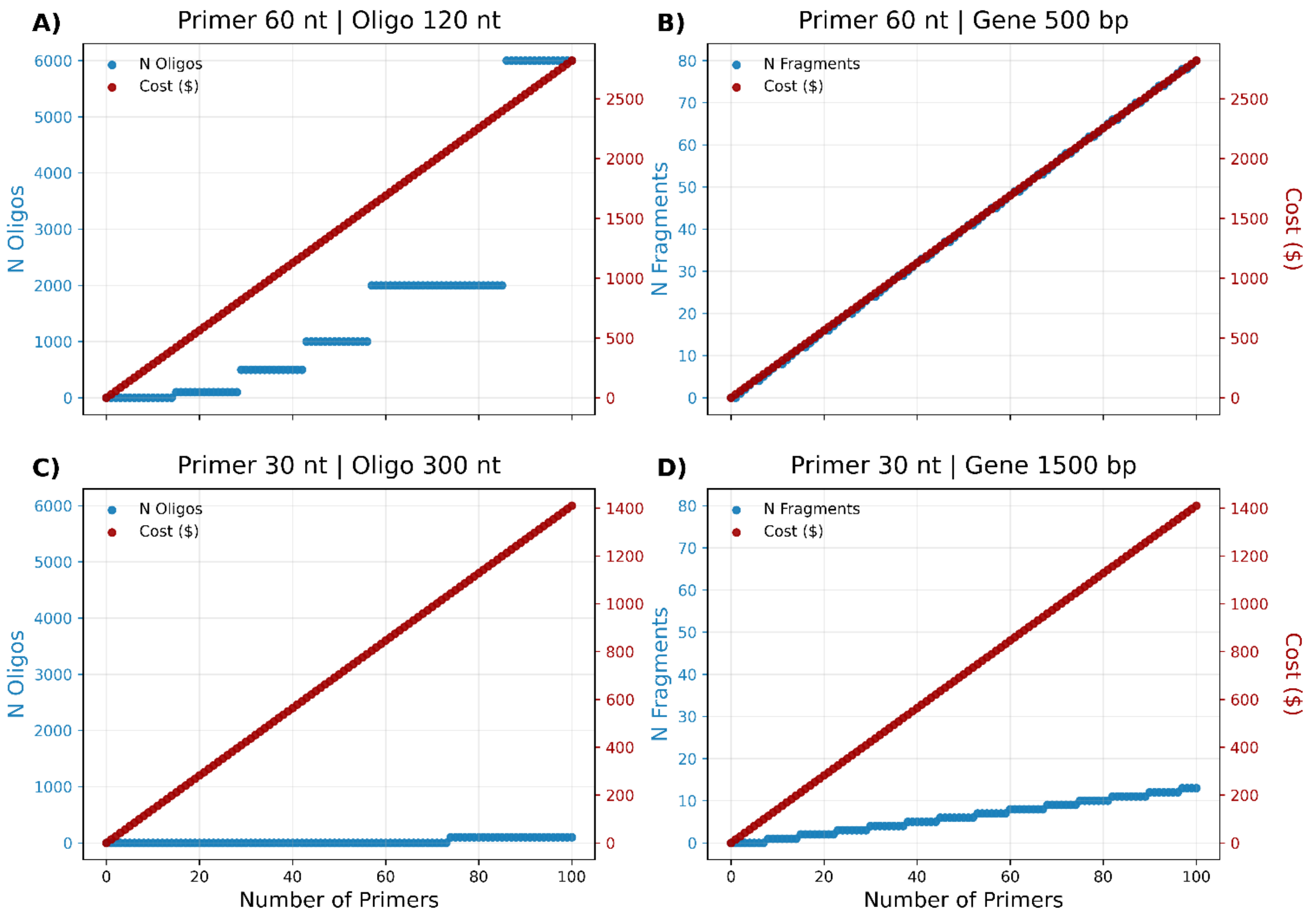

**Figure 2.** Cost for N primers, and the maximum number of oligos in an oligo pool (left column) or gene fragments (right column) that could be ordered at that cost. The x-axis is always the number of primers. The secondary y-axis is the cost of ordering that number of primers. The primary y-axis is the number of oligos or genes that could be ordered for that cost. The top row assumes 60-nt primers, 120-nt oligo pool oligos (A), and 500-nt gene fragments (B). The bottom row assumes 30-nt primers, 300-nt oligo pool oligos (C), and 1500-nt gene fragments (D). Primer costs were calculated using pricing from IDT, and oligo pool and gene fragment costs were calculated using pricing from Twist Bioscience, both as of May 28, 2026.

As currently employed, for MLDE to be competitive with traditional DE, it must be able to do more with less to offset its added per-cycle costs; it must be able to find a sufficiently fit variant faster (e.g., with fewer laboratory cycles) and/or using less resources (e.g., fewer variants screened). But can it do this? Sometimes, absolutely. MLDE can operate with fewer variants than DE, so when the cost of screening dominates and it becomes infeasible to evaluate even site-saturation libraries, MLDE can be effective for prioritizing those few variants that can be screened within budget.[38,39] Consistently, though, it has been observed that current “state-of-the-art” MLDE approaches to fitness prediction show comparable utility to simple linear regression, an ML modeling approach that by definition gives equivalent predictions to—and so is subject to the same limitations of—the traditional strategy of naïvely recombining beneficial mutations.[16–21,40,41] Thus, if the cost of screening is sufficiently low to gather enough variants for naïve recombination, for the same screening budget, traditional DE and MLDE will tend to yield similar results. For most screens, the cost is low enough to enable this. It follows that except in cases where the fitness landscape in question is known to be challenging to model with a linear model,[19,42] as the option with lower DNA synthesis costs, traditional DE will be the more appropriate option, identifying similarly fit variants as MLDE for lower cost overall.

## Scaling Is Not the Solution

Much of the last five years of work in the broader space of ML for protein fitness prediction has focused on strategies that train ever-larger models on ever-larger sources of unlabeled data, then use the resulting biological foundation models as a vehicle for transfer learning in downstream tasks. As highlighted in the original perspective,[15] the underlying assumption of such unsupervised and semi-supervised strategies is that a protein language (or similar) model trained on a sufficiently large corpus of unlabeled natural protein sequence data will learn the set of "biophysical rules" that govern proteins' abilities to be produced and carry out functions, and that by extracting these rules, the amount of data required for downstream tasks can be reduced.

Then as now, I agree with the sentiment of this strategy: As just discussed, MLDE becomes far more applicable as its screening requirements drop, so learning from readily available unlabeled data to reduce the need for collecting costly labeled data is appealing. Unfortunately, however, unlike with large language models, which do appear to exhibit so-called "emergent capabilities" as scales increase, growing evidence suggests that scaling does not yield improved capabilities for MLDE, at least with current pretraining strategies.[23–25,40,41,43–46] And perhaps this should not be surprising.

While it is a nice abstraction to describe protein language models as learning biophysical rules, that is not what they are doing. What they actually learn is better described as "sequence statistics": correlates of biophysical rules such as signatures of stability, contact patterns, and conserved motifs that are downstream of physics and chemistry but are not the physics and chemistry themselves.[43,47–51] Given this, it is unreasonable to expect that these models would ever be sufficient for all protein engineering campaigns, particularly when engineering toward functions not found in nature, which by definition are unlikely to be reflected by the sequence statistics of natural proteins, and which are also often the focus of DE campaigns. Indeed, while unsupervised and semi-supervised MLDE methodologies have seen success when applied to known functions (i.e., functions represented in the training corpus), the same cannot be said for proteins with complex mechanisms of action or "non-natural" definitions of fitness, strongly suggesting that chemistry and physics are not inherent to what an unsupervised model learns.[50] This notwithstanding, a protein's "fitness" is contextual, with different definitions of fitness frequently opposing one another.[52] Models trained on unlabeled data alone do not have the context to determine, for example, whether to prioritize an enzyme's stability or its catalytic efficiency, should a proposed mutation be to the benefit of one definition of fitness at the expense of the other.

This is not at all to say that I believe efforts to learn from unlabeled data have been a waste, just that the prevailing strategies for using the information they learn are often inappropriate for MLDE. Indeed, as discussed in the next and final section, I believe that the generative models trained from such data can be a critical component of a "distribution-focused" approach to MLDE that achieves cost parity with traditional strategies.

## Distribution-Focused MLDE as a Path Forward

MLDE is currently a variant-focused strategy; it aims to answer the question, "which N *individual* variants do I order?", and it is the answer to this question that determines the makeup of the oligo pools or gene arrays discussed earlier. Traditional approaches to DE, however, do not take this approach. The protein engineer building a library via error-prone PCR is under no illusion that they will be screening every variant they construct. Instead, they toggle various library synthesis parameters that they predict will (1) maximize their chance of identifying an improved variant from the library while (2) keeping the overall screening burden within whatever budget they can afford. Indeed, these types of decisions—where an engineer aims to maximize the *probability* of finding an acceptable variant *given* experimental constraints—abound in protein engineering workflows: Active sites tend to be targeted for site-saturation mutagenesis before surface residues; error-prone mutation rates are tuned to maximize diversity while minimizing the number of non-functional variants; single mutants tend to be prioritized for screening over higher-order mutants, etc.

While it may not be typically framed as such, traditional DE is inherently a game of probability, and MLDE should be adapted to follow suit. Instead of focusing on predicting or generating specific variants, it should focus on variant *distributions*. It should answer the question, "what parameters of a synthesis-feasible distribution (e.g., degenerate oligo makeup, recombination block boundaries, error rates for error-prone PCR, etc.) maximize the probability of finding at least one sufficiently fit variant over the course of my engineering campaign?" Or more simply, "how do we minimize synthesis cost while maximizing the probability that we will draw a sufficiently fit variant from one of our libraries?" This small change in framing has large practical implications, as it directly aligns the computational objective with real-world constraints by including library synthesis and other experimental costs as explicit optimization targets.

Excellent examples of this "distribution-focused" approach to MLDE already exist via the works of Weinstein et al. and Sussex et al., who similarly observed that the standard way of testing a generative sequence model—sampling sequences from it and then synthesizing each one individually—is fundamentally mismatched with the economics of the experiment.[53–55] In their approaches—"variational synthesis" and "policy gradients for library design" (PGLD), respectively—rather than asking which individual sequences to build, they write down a probabilistic model of the sequences that a given stochastic synthesis protocol will actually produce, parameterized directly according to the laboratory parameters the experimentalist can tune. They then optimize those parameters to minimize the divergence between the protocol's output distribution and the target distribution (e.g., that defined by a generative model), so that simply running the protocol in the laboratory draws approximate samples from the model. The trade is one of accuracy for scale: the synthesized distribution will not match the target exactly, but the resulting bias is more than offset by the orders-of-magnitude increase in the number of sequences that can be made and screened, all for the (significantly lower) cost of a stochastic library instead of a targeted synthesis library. It is, in other words, a concrete demonstration that aligning the computational objective with what can be synthesized is worth more than chasing the "best" individual designs.

And this is where unsupervised and semi-supervised learning comes back into play: Distribution-focused methods like variational synthesis and PGLD are complementary to the broader efforts in generative biology and protein fitness prediction made over the last five years. Increasingly, researchers treat the learned probability distributions underpinning unsupervised generative models as Bayesian priors ($p(s)$) over natural sequence space. By drawing (i.e., generating) from these priors and evaluating the resultant sequences in the laboratory, fitness prediction models can be trained and then used to produce a posterior distribution ($p(s\,|f_1)$). This posterior is then used to generate more samples, and the process repeats, with each iteration (ideally) resulting in a model ($p(s\,|f_1 \dots f_n)$) that has a higher probability of generating variants that are "fit for purpose" (Fig 3).[27,56–59] This approach neatly takes advantage of both unlabeled data (capturing information from sequence statistics) and labeled data (capturing experiment-specific "fitness" information) while also accounting for the iterations inherent to laboratory work. It is already well-aligned with laboratory practices, but could be made even more so by conditioning generation of sets of sequences not only on laboratory data, but also their buildability ($p(s\,|f_1 \dots f_n, b)$). In other words, the framework is already there; MLDE

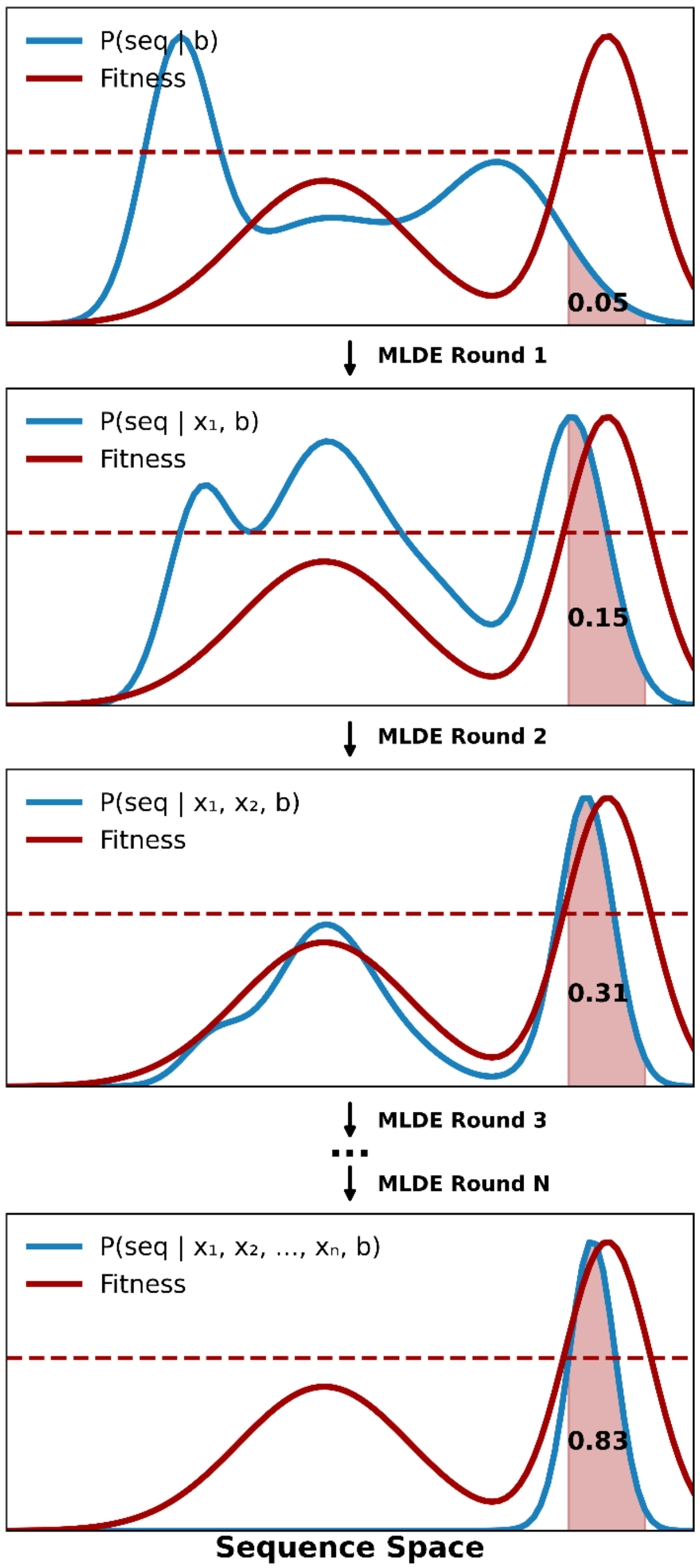

**Figure 3.** Active learning with generative models can be thought of as progressively updating/collapsing a probability distribution to maximize the chance that a sufficiently fit protein sequence will be drawn from it.

researchers are thinking probabilistically; we just need to find a way to add one more conditional.

## Conclusion

Prior to AlphaFold, obtaining a protein structure involved a unique skillset, specialized equipment, and weeks to months of laboratory time, all with no guarantee of success. By contrast, using AlphaFold, anyone with access to a computer can obtain structures in a matter of seconds to minutes. And, while these predicted structures certainly have their limitations—AlphaFold is not perfect—the dramatic reductions in cost and difficulty of obtaining them more than compensate for any of their deficiencies.

Similarly, prior to generative AI, de novo design of a fit-for-purpose protein, redesign of an existing protein, and numerous other protein-design-related manipulation and engineering tasks were possible using tools like Rosetta, but success rates were low.[60] By contrast, generative protein design tools have considerably higher success rates, and are just as accessible if not more than the classical tools they have largely come to replace. With no-to-minimal additional overhead and a higher success rate, it is no surprise that they have risen to such immediate prominence.

Now compare these examples to MLDE. Prior to MLDE, DE certainly had its limitations, but it was already an extremely effective technology.[61] Unlike with protein structure prediction or protein design, then, the bar for a truly transformative MLDE technology is much higher. Possibly, when such a technology arrives, it will come in the form of an extremely capable generative or—as suggested in my earlier perspective—fitness-prediction model; though, such a model would need capabilities far beyond the current state of the art to compete with DE in an applied setting, where the cost of obtaining an outcome matters just as much as the outcome itself.

More than likely, MLDE's "AlphaFold moment" will come from a technology that optimizes for all variables relevant to protein fitness optimization: fitness, cost, and time. Within this framing, we do not need bigger models; we do not need more data; we need to recognize and value the environment in which MLDE is employed. Performance on fitness benchmarks cannot be the priority; the laboratory cannot be an afterthought; and unless our priorities change, I fear the picture five years from now may not be too different.

## Acknowledgments

Thank you to Jinbei Li and Adam Meyer for encouraging me to write this article, early feedback, assistance developing the initial outline, and critical reading of early drafts. Thank you to Ella Watkins-Dulaney, Shilong Gao, Zachary Wu, Kevin Yang, and Céline Marquet for critical reading and feedback.

## References

1. Guo, D. *et al.* DeepSeek-R1 incentivizes reasoning in LLMs through reinforcement learning. *Nature* **645**, 633–638 (2025).
2. OpenAI *et al.* GPT-4 technical report. Preprint at https://doi.org/10.48550/arXiv.2303.08774 (2024).
3. Maslej, N. *et al.* Artificial intelligence index report 2025. Preprint at https://doi.org/10.48550/ARXIV.2504.07139 (2025).
4. Let 2026 be the year the world comes together for AI safety. *Nature* **649**, 7–7 (2026).

5. Brauchler, D. *Analyzing AI Application Threat Models*. https://www.nccgroup.com/research-blog/analyzing-ai-application-threat-models/ (2024).

6. *Frontier AI Trends Report*. https://www.aisi.gov.uk/frontier-ai-trends-report.

7. Mitch, I. *et al. Governance Approaches to Securing Frontier AI*. https://www.rand.org/pubs/research_reports/RRA4159-1.html (2025).

8. Dauparas, J. *et al.* Robust deep learning–based protein sequence design using ProteinMPNN. *Science* **378**, 49–56 (2022).

9. Watson, J. L. *et al.* De novo design of protein structure and function with RFdiffusion. *Nature* **620**, 1089–1100 (2023).

10. Jumper, J. *et al.* Highly accurate protein structure prediction with AlphaFold. *Nature* **596**, 583–589 (2021).

11. *2026 Biotech AI Report*. https://www.benchling.com/biotech-ai-report-2026 (2026).

12. OpenAI. Biodefense in the intelligence age. https://openai.com/index/biodefense-in-the-intelligence-age/ (2026).

13. Brent, R. & McKelvey, G. Contemporary Foundation AI Models Increase Biological Weapons Risk.

14. Webster, T. *et al. Global Risk Index for AI-Enabled Biological Tools (Public Report)*. https://www.longtermresilience.org/reports/global-risk-index-for-ai-enabled-biological-tools/ (2025) doi:10.71172/wjyw-6dyc.

15. Wittmann, B. J., Johnston, K. E., Wu, Z. & Arnold, F. H. Advances in machine learning for directed evolution. *Curr. Opin. Struct. Biol.* **69**, 11–18 (2021).

16. Dallago, C. *et al.* FLIP: Benchmark tasks in fitness landscape inference for proteins. Preprint at https://doi.org/10.1101/2021.11.09.467890 (2021).

17. Didi, K. *et al.* FLIP2: Expanding Protein Fitness Landscape Benchmarks for Real-World Machine Learning Applications. Preprint at https://doi.org/10.64898/2026.02.23.707496 (2026).

18. Hsu, C., Nisonoff, H., Fannjiang, C. & Listgarten, J. Learning protein fitness models from evolutionary and assay-labeled data. *Nat. Biotechnol.* **40**, 1114–1122 (2022).

19. Wittmann, B. J., Yue, Y. & Arnold, F. H. Informed training set design enables efficient machine learning-assisted directed protein evolution. *Cell Syst.* **12**, 1026-1045.e7 (2021).

20. Johnston, K. E. *et al.* A combinatorially complete epistatic fitness landscape in an enzyme active site. *Proc. Natl. Acad. Sci.* **121**, e2400439121 (2024).

21. Li, F.-Z. *et al.* Evaluation of machine learning-assisted directed evolution across diverse combinatorial landscapes. *Cell Syst.* **16**, 101387 (2025).

22. Bhatnagar, A. *et al.* Scaling Unlocks Broader Generation and Deeper Functional Understanding of Proteins. Preprint at https://doi.org/10.1101/2025.04.15.649055 (2025).

23. Spinner, A., DeBenedictis, E. & Hudson, C. M. Scaling and Data Saturation in Protein Language Models. Preprint at https://doi.org/10.48550/ARXIV.2507.22210 (2025).

24. Fournier, Q. *et al.* Protein Language Models: Is Scaling Necessary? Preprint at https://doi.org/10.1101/2024.09.23.614603 (2024).

25. Li, F.-Z., Amini, A. P., Yue, Y., Yang, K. K. & Lu, A. X. Feature Reuse and Scaling: Understanding Transfer Learning with Protein Language Models. Preprint at https://doi.org/10.1101/2024.02.05.578959 (2024).

26. Yang, K. K. *et al.* The Dayhoff Atlas: scaling sequence diversity for improved protein generation. Preprint at https://doi.org/10.1101/2025.07.21.665991 (2025).

27. Zhu, Y. *et al.* Generative AI for Controllable Protein Sequence Design: A Survey. Preprint at https://doi.org/10.48550/ARXIV.2402.10516 (2024).

28. Morris, M. R. Scientists' Perspectives on the Potential for Generative AI in their Fields. Preprint at https://doi.org/10.48550/ARXIV.2304.01420 (2023).

29. Moldwin, A. & Shehu, A. Foundation models for AI-enabled biological design. Preprint at https://doi.org/10.48550/ARXIV.2505.11610 (2025).

30. Baek, M. *et al.* Accurate prediction of protein structures and interactions using a three-track neural network. *Science* **373**, 871–876 (2021).

31. Notin, P. *et al.* ProteinGym: Large-Scale Benchmarks for Protein Fitness Prediction and Design. in *Advances in Neural Information Processing Systems 36* doi:10.52202/075280-2810.

32. McLellan, J. R. *et al.* Functional profiling of thousands of sequence-diverse protease homologs with GROQ-seq. Preprint at https://doi.org/10.64898/2026.04.30.721934 (2026).

33. Fowler, D. M. & Fields, S. Deep mutational scanning: a new style of protein science. *Nat. Methods* **11**, 801–807 (2014).

34. Li, J. *et al.* MillionFull enables low-cost, massive, full-length enzyme sequence–fitness data collection for machine learning–guided enzyme engineering. Preprint at https://doi.org/10.1101/2025.10.24.684421 (2025).

35. Long, Y. *et al.* LevSeq: Rapid Generation of Sequence-Function Data for Directed Evolution and Machine Learning. *ACS Synth. Biol.* **14**, 230–238 (2025).

36. Wittmann, B. J., Johnston, K. E., Almhjell, P. J. & Arnold, F. H. evSeq: Cost-Effective Amplicon Sequencing of Every Variant in a Protein Library. *ACS Synth. Biol.* **11**, 1313–1324 (2022).

37. Yang, J. *et al.* DeCOIL: Optimization of Degenerate Codon Libraries for Machine Learning-Assisted Protein Engineering. *ACS Synth. Biol.* **12**, 2444–2454 (2023).

38. Biswas, S., Khimulya, G., Alley, E. C., Esvelt, K. M. & Church, G. M. Low-N protein engineering with data-efficient deep learning. *Nat. Methods* **18**, 389–396 (2021).

39. Bedbrook, C. N. *et al.* Machine learning-guided channelrhodopsin engineering enables minimally invasive optogenetics. *Nat. Methods* **16**, 1176–1184 (2019).

40. Talpir, I. & Fleishman, S. J. Simple baselines rival protein language models in mutation-dense design of function tasks. Preprint at https://doi.org/10.64898/2026.05.01.722313 (2026).

41. Kolchina, A., Dubanevics, I., Kondrashov, F. A. & Kalinina, O. V. Beyond additivity: zero-shot methods cannot predict impact of epistasis on protein properties and function. Preprint at https://doi.org/10.64898/2026.02.17.706292 (2026).

42. Wu, Z., Kan, S. B. J., Lewis, R. D., Wittmann, B. J. & Arnold, F. H. Machine learning-assisted directed protein evolution with combinatorial libraries. *Proc. Natl. Acad. Sci.* **116**, 8852–8858 (2019).

43. Ursu, E. *et al.* Training data composition determines machine learning generalization and biological rule discovery. *Nat. Mach. Intell.* **7**, 1206–1219 (2025).

44. Hou, C., Liu, D., Zafar, A. & Shen, Y. Understanding language model scaling for protein fitness prediction. *Nat. Comput. Sci.* 1–19 (2026) doi:10.1038/s43588-026-01010-z.

45. Bubeck, S. *et al.* Sparks of artificial general intelligence: early experiments with GPT-4. Preprint at https://doi.org/10.48550/ARXIV.2303.12712 (2023).

46. Spinner, A., DeBenedictis, E. & Hudson, C. M. Scaling and data saturation in protein language models. Preprint at

https://doi.org/10.48550/ARXIV.2507.22210 (2025).

47. Shaw, A. *et al.* Removing bias in sequence models of protein fitness. Preprint at https://doi.org/10.1101/2023.09.28.560044 (2023).

48. Zhang, Z. *et al.* Protein language models learn evolutionary statistics of interacting sequence motifs. *Proc. Natl. Acad. Sci.* **121**, e2406285121 (2024).

49. Ding, F. & Steinhardt, J. Protein language models are biased by unequal sequence sampling across the tree of life. Preprint at https://doi.org/10.1101/2024.03.07.584001 (2024).

50. Berry, S. P., Gaudet, R. & Marks, D. S. Differences between protein fitness models can be used to design variants of altered specificity. Preprint at https://doi.org/10.64898/2026.06.10.731299 (2026).

51. Yang, J., Li, F.-Z., Long, Y. & Arnold, F. H. Illuminating the universe of enzyme catalysis in the era of artificial intelligence. *Cell Syst.* **17**, 101372 (2026).

52. Pugh, C. W. J., Nuñez-Valencia, P. G., Dias, M. & Frazer, J. From Likelihood to Fitness: Improving Variant Effect Prediction in Protein and Genome Language Models. Preprint at https://doi.org/10.1101/2025.05.20.655154 (2025).

53. Weinstein, E. N. *et al.* Manufacturing-aware generative models enable petascale synthesis of designed DNA. *Nat. Biotechnol.* https://doi.org/10.1038/s41587-026-03020-8 (2026) doi:10.1038/s41587-026-03020-8.

54. Weinstein, E. N. *et al.* Optimal Design of Stochastic DNA Synthesis Protocols based on Generative Sequence Models. Preprint at https://doi.org/10.1101/2021.10.28.466307 (2021).

55. Sussex, S. *et al.* Breaking the synthesis barrier for AI-designed DNA libraries. Preprint at https://doi.org/10.64898/2026.07.07.736931 (2026).

56. Lisanza, S. L. *et al.* Multistate and functional protein design using RoseTTAFold sequence space diffusion. *Nat. Biotechnol.* **43**, 1288–1298 (2025).

57. Xiong, J. *et al.* ProteinGuide: On-the-fly property guidance for protein sequence generative models. Preprint at https://doi.org/10.48550/ARXIV.2505.04823 (2025).

58. Yang, J. *et al.* Steering generative models with experimental data for protein fitness optimization. Preprint at https://doi.org/10.48550/ARXIV.2505.15093 (2025).

59. Yuan, S. C. *et al.* Conditioning protein language models using high-throughput sequence-fitness data collection. in (3/12026).

60. Leman, J. K. *et al.* Macromolecular modeling and design in Rosetta: recent methods and frameworks. *Nat. Methods* **17**, 665–680 (2020).

61. *Scientific Background on the Nobel Prize in Chemistry 2018*. https://www.nobelprize.org/prizes/chemistry/2018/advanced-information/ (2018).